\documentclass[aps,prl,reprint,superscriptaddress,nobibnotes,floatfix]{revtex4-2}
\usepackage{graphicx}
\usepackage{dcolumn}
\usepackage{blindtext}
\usepackage{lineno}
\usepackage{amsmath}
\usepackage{amssymb}
\usepackage{bbold}
\usepackage{float}
\usepackage{newtxtext,newtxmath}
\usepackage[utf8]{inputenc}
\usepackage[T1]{fontenc}
\usepackage{mathtools}

\usepackage{hyperref}
\usepackage{braket}
\usepackage{multirow}
\usepackage{siunitx}
\usepackage[dvipsnames]{xcolor}

\newcommand\identity{1\kern-0.25em\text{l}}

\newcommand*\Diff[1]{\mathop{}\!\mathrm{d^#1}}

\colorlet{mylinkcolor}{RoyalPurple}
\colorlet{mycitecolor}{RoyalPurple}
\colorlet{myurlcolor}{RoyalPurple}
\hypersetup{
	linkcolor  = mylinkcolor,
	citecolor  = mycitecolor,
	urlcolor   = myurlcolor,
	colorlinks = true,
	breaklinks = true
}

\newcommand{\subfig}[1]{(#1)}
\newcommand{\iu}{{i\mkern1mu}}

\newcommand{\ketbra}[2]{| #1 \rangle\langle #2 |}
\newcommand{\proj}[1]{| #1 \rangle \langle #1|}
\begin{document}
\title{Microwave-field quantum metrology with inherent robustness against detection losses enabled by Rydberg interactions}

\author{Stanisław Kurzyna}
\thanks{Equal contributions}
\affiliation{Centre for Quantum Optical Technologies, Centre of New Technologies, University of Warsaw, Banacha 2c, 02-097 Warsaw, Poland}
\affiliation{Faculty of Physics, University of Warsaw, Pasteura 5, 02-093 Warsaw, Poland}
\author{Bartosz Niewelt}
\thanks{Equal contributions}
\affiliation{Centre for Quantum Optical Technologies, Centre of New Technologies, University of Warsaw, Banacha 2c, 02-097 Warsaw, Poland}
\affiliation{Faculty of Physics, University of Warsaw, Pasteura 5, 02-093 Warsaw, Poland}
\author{Mateusz Mazelanik}
\affiliation{Centre for Quantum Optical Technologies, Centre of New Technologies, University of Warsaw, Banacha 2c, 02-097 Warsaw, Poland}

\author{Wojciech Wasilewski}
\affiliation{Centre for Quantum Optical Technologies, Centre of New Technologies, University of Warsaw, Banacha 2c, 02-097 Warsaw, Poland}
\affiliation{Faculty of Physics, University of Warsaw, Pasteura 5, 02-093 Warsaw, Poland}
\author{Rafał Demkowicz-Dobrzański}
\affiliation{Faculty of Physics, University of Warsaw, Pasteura 5, 02-093 Warsaw, Poland}
\author{Michał Parniak}
\email{michal.parniak@uw.edu.pl}
\affiliation{Centre for Quantum Optical Technologies, Centre of New Technologies, University of Warsaw, Banacha 2c, 02-097 Warsaw, Poland}
\affiliation{Faculty of Physics, University of Warsaw, Pasteura 5, 02-093 Warsaw, Poland}

\begin{abstract}
Quantum sensing and metrology present one of the most promising near-term applications in the field of quantum technologies, with quantum sensors enabling unprecedented precision in measurements of electric, magnetic or gravitational fields and displacements. Experimental loss at the detection stage remains one of the key obstacles to achieving a truly quantum advantage in many practical scenarios. Here, we combine the capabilities of Rydberg atoms to both sense external fields and be used for quantum information processing, thereby largely overcoming the issue of detection losses. While utilising the large dipole moments of Rydberg atoms in an ensemble to achieve a $\SI{39}{\nV\per\cm \hertz\tothe{-1/2}}$ sensitivity, we employ inter-atomic dipolar interactions to take advantage of an error-prevention protocol that protects information against conventional losses at the detection stage. Counterintuitively, the protocol's idea is based on introducing an additional non-linear, lossy quantum channel, which results in a 3.3-fold enhancement of Fisher information. The presented results pave the way for broader adoption of quantum-information-inspired enhancements enabled by intrinsic interactions present in a sensor system, and more broadly  in practical quantum metrology and communication, without the need for a general-purpose quantum computer.
\end{abstract}

\maketitle

\section{Introduction}
Rydberg atoms currently find exciting applications both in quantum computing or simulation and in quantum metrology. Those atomic systems provide a controllable platform to engineer atomic interactions \cite{Li2016,PhysRevLett.108.013002,Zhao2023} enabling reliable quantum computation \cite{Bluvstein2024, Cohen2021, Saffman2016, PhysRevLett.127.063604}, quantum simulations \cite{Weimer2010, Scholl2021, Browaeys2020}, and high-fidelity entanglement generation for quantum networks \cite{Madjarov2020, Li2013, Saffman2002,PhysRevLett.108.030501}.  Recent developments of Rydberg-assisted atomic sensors have demonstrated their exciting new role in imaging of electromagnetic fields \cite{Wade2017}, quantum transduction \cite{Borówka2024,Kumar2023} and microwave sensing with high sensitivity \cite{Kumar:17}.

For metrological applications, it is an essential advantage to use as many sensing particles as possible rather than single atoms. 
In particular, one may employ collective atomic states in atomic ensembles to not only use many atoms for the sensing task itself, but also enhance the coupling to light and thus the read-out rate of the sensors \cite{PhysRevA.109.052615, Shaw2024, Robinson2024,maxwell_storage_2013}. In synergy with a large dipole moment of Rydberg-excited atoms, those sensors hold a promise to be a majorly disruptive technology. Atomic qubits made with collective states \cite{PhysRevLett.127.063604} show promising applications also in quantum computing and quantum networking, enabling Rydberg-mediated photon-photon interactions \cite{Tiarks2019} or atom-ensemble entanglement \cite{PhysRevLett.123.140504}. The lifetime of those qubits has also been extended significantly with recent advances \cite{Kurzyna2024longlivedcollective, PhysRevLett.134.053604}.

The combination of metrological and quantum-computational applications of Rydberg atoms presents a challenge in integrating these concepts, leveraging the versatility of quantum information processing to enhance metrological performance. Another essential idea is to attempt to use error correction enabled by quantum operations to improve metrology protocols \cite{Kessler2014}. These ideas sparked a plethora of efforts to design new protocols \cite{Arrad2014-vw, Zhou2018, Dur2014-wp, Layden2019-wq, Lu2015-fo, Hainzer2024-it, Matsuzaki2017-pe, Kwon2025}, especially in the context of NV-center and ion-trap systems. Most efforts in this direction for Rydberg-atom systems so far have been centered around atomic-optical clocks \cite{PhysRevApplied.17.064050, PRXQuantum.4.020322}. Overall, the theoretical advances have so far exceeded the experimental capabilities, hindering the practical application of error-corrected quantum metrology.

Here, we show that, by proper use of intrinsic interactions between Rydberg atoms, it is possible to enhance the performance in the practical task of microwave sensing, thanks to increased robustness against measurement noise. This phenomenon may be understood in terms of an idea of `premeasurement processing' \cite{PRXQuantum.4.040305,Len2022-vt}, which refers to the fact that in presence of imperfect measurements, a properly tailored additional quantum operation performed on the state, before it is subject to measurement, may make it less susceptible to measurement noise and allow better extraction of encoded information. This effect has been exploited in a number of sensing experiments \cite{Hosten2016, Linnemann2016, Li2023}, but have never been demonstrated in Rydberg atomic systems.

 The optimal estimation in real conditions is captured by the Cramér-Rao bound \cite{Fisher_book}, which ties the amount of Fisher information (FI) with the maximum possible precision of the measurement. In case of estimation of a parameter encoded in a quantum state $\rho_\theta$, the ultimate precision limit is set by the quantum Fisher information (QFI) $\mathcal{F}_Q\left(\rho_\theta \right)$ \cite{Helstrom1976, Braunstein1994}:
\begin{equation}
\label{eq:qcrb}
\Delta \theta \geq 1/\sqrt{\mathcal{F}_Q\left(\rho_\theta\right)},
\end{equation}
where optimisation over all measurement schemes is taken into account. In typical situations, involving independent sensing probes (or independent mode excitations) where detection losses act independently on each sensing particle, 
the obtainable information is reduced to $\eta\, \mathcal{F}_Q$, where $\eta$ is the detection efficiency.

Remarkably, in the system we consider, \emph{additional non-linear losses} present in the  setup make the sensing particles more robust to subsequent detection losses, and effectively increase the FI compared to the system without the added losses.

We identified the influence of the intrinsic Rydberg interatomic interaction, verifying the theoretical models of the interacting particles and benchmarking the sensitivity to weak microwave fields enabled by the huge dipole moments.
The collective states allow excellent coupling to light, and the large size of the atomic ensemble enables the use of many sensor particles in each experimental shot. We report that the error preventing operation is built-in into the Rydberg sensor particles themselves, hence it takes advantage of various aspects of intricate Rydberg physics simultaneously. We also take inspiration from the progress on lifetime prolongation methods \cite{Kurzyna2024longlivedcollective} and use a novel protocol that stops the thermal decoherence caused by atomic motion, making long-lasting interactions possible. 
With this, we achieve both excellent and quantum-enhanced sensitivity to an external microwave field, showing a direct application of error prevention in quantum metrology \cite{Takagi2022}.

Furthermore, using novel tools designed to identify the optimal quantum metrological protocols in the presence of noise, we show that the mechanism observed is fundamentally optimal in the simplest non-trivial case of two sensing particles, proving no other quantum protocol, no matter how general, could offer better performance given the level of detection losses present.

\section{Results}

\subsection{Theoretical model}
\begin{figure}
    \centering
    \includegraphics[width=1\linewidth]{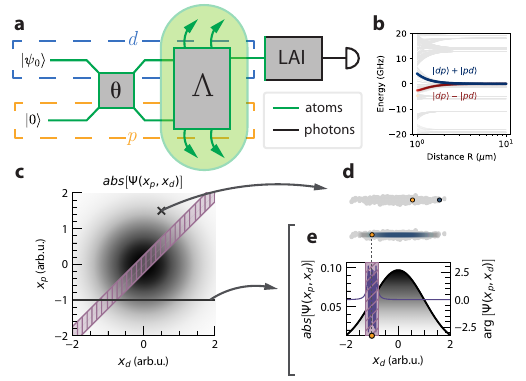}
    \caption{
    \subfig{a} Ideational scheme of the presented experiment. Initially, photons are stored in the atomic ensemble as a collective Rydberg excitation $\ket{\psi_0}$ in the mode $d$. Afterwards, the spinwaves created that way are placed in the superposition between modes $d$ and $p$ representing the two different Rydberg atomic levels with rotation $\theta$. This allows for the dipolar Rydberg-Rydberg interactions, denoted as $\Lambda$, to scatter the collective excitations in different modes (depicted with green area). LAI (light-atoms interface) is applied to convert the stored excitations to photons, allowing for the detection with the overall efficiency $\eta$.
    \subfig{b} Pair interaction strength for Rydberg states $\ket{d} = \ket{49^2D_{5/2}\;m_J=5/2}$ and $\ket{p} = \ket{50^2P_{3/2}\;m_J=3/2}$. Energy of symmetric superposition $\ket{dp}+\ket{pd}$ is denoted with a blue line, whereas the red line represents the energy of anti-symmetric superposition.
    \subfig{c} Two-excitation spinwave wavefunction $\Psi(x_p,x_d)$ with one excitation in $\ket{p}$ and position $x_p$ and the other in state $\ket{d}$ and position $x_d$. Depending on the distance between the excitations, a part of the wavefunction will acquire additional phase due to the pair interactions (depicted with a hatched region).
    \subfig{d} Elements of the two-excitation collective superposition. Blue and orange circles represent atoms in state $\ket{d}$ and $\ket{p}$, respectively. The grey circles in the background represent the rest of the atoms in the ensemble.
    \subfig{d} Elements of the two-excitation collective superposition with fixed position of the atom in state $\ket{p}$. The blurred blue region represents an atom in state $\ket{d}$ in spatial superposition. Below, the plot shows the absolute value and phase of a slice of the wavefunction. The region around the fixed atom in state $\ket{p}$ acquires an additional phase.
    }
    \label{fig:blend}
\end{figure}

To identify the influence of the intrinsic Rydberg interactions on the sensitivity of the microwave field detection, we utilised an ultra-cold atomic ensemble capable of storing collective Rydberg excitations. 
Atoms in the same Rydberg states exhibit a short-range van der Waals (vdW) type interaction scaling with distance as $R^{-6}$. This manifests in their remarkable properties crucial for the implementation of quantum gates \cite{Evered2023} and single photon generation \cite{Shi2022}. However, once two different Rydberg excitations are introduced, due to the dipolar coupling, they interact considerably stronger at large distances with a potential scaling with distance as $R^{-3}$. The energy shifts resulting from interactions \cite{Weber2017} between two atoms in states $\ket{d} = \ket{49^2D_{5/2}\;m_J=5/2}$ and $\ket{p} = \ket{49^2D_{5/2}\;m_J=5/2}$ are displayed in Fig.~\ref{fig:blend}\subfig{b}. In the experiment, the timing, internal states and the number of excitations are chosen in such a way that the dominant type of interactions is dipolar (see \hyperref[sec:methods]{Methods}).

The ideational scheme of the experiment is presented in the Fig.~\ref{fig:blend}\subfig{a}. Using the light-atoms interface \cite{mazelanik_coherent_2019} (LAI), we prepare the initial state of the ensemble of atoms in a collective superposition (spinwave) of a Rydberg state $\ket{d}$ (spinwave in mode $d$), denoted as $\ket{\psi_0}$.
If we now consider only a single excitation in mode $d$, the state of the spinwave would be $\ket{\psi_0} = \ket{1}_{d,\boldsymbol{k}} = \hat{d}_{\boldsymbol{k}}^{\dagger} \ket{g_1 \ldots g_N} = \frac{1}{\sqrt{N}}\sum_{m}\exp{(- \iu\, \boldsymbol{k} \cdot \boldsymbol{r}_m)} \ket{g_1\ldots d_m \ldots g_N}$, where $\boldsymbol{k}$ and $\boldsymbol{r}_m$ are respectively the wavevector and the position of the $m$-th atom in the ensemble. Initially, we prepare the system in a flat spinwave state ($\boldsymbol{k} = 0$).
Throughout the work, as a sensor particle, we understand a single collective Rydberg excitation, which corresponds to a single delocalized and excited atom.

We also introduce to the system another Rydberg excitation $\ket{p}$. It follows that there is an analogous spinwave mode $p$, created by an operator $\hat{p}_{\boldsymbol{k}}^{\dagger}$, similarly to mode $d$. One can drive Rabi oscillations between the two Rydberg states (and so the two modes) by applying a microwave (MW) field. By doing so, the angle of oscillation $\theta$, which holds information about the amplitude of the electric field, is encoded onto the atomic superposition. This operation corresponds to block $\theta$ in the ideational scheme.

Under the dipolar interactions, the system will evolve with a complex many-body Hamiltonian:
\begin{equation}
    \hat{H}=\sum_{\mu<\nu} \hat{\sigma}_{dp}^\mu\hat{\sigma}_{pd}^\nu V^{\mu\nu}_{dip} + h.c.
\end{equation}
The energy of two resonantly interacting Rydberg excitations is equal to the energy of two dipoles separated by a distance $R$, oriented along the propagation axis $z$:
\begin{equation}
V_{dip}^{\mu  \nu } (R, \vartheta)= \frac{C_3}{ R^3}\frac{3 \cos (2 \vartheta )+1}{4}
\end{equation}
With $\vartheta$ being the angle between the axis connecting the two dipoles and the $z$ axis.
The long-range character of the Hamiltonian does not allow for any simple treatment. Still, we may adopt a simplified model assuming that we are in fact only interested in the flat spin-wave state ($\boldsymbol{k}=0$). Furthermore, we only consider pair, or two-particle, effects and ignore any higher-order terms. The dipolar interactions correspond to block $\Lambda$ in the ideational scheme.

Now, let us consider the state of two delocalized excitations in both modes of the spinwave (the state is $\ket{1}_{d,\boldsymbol{k}=0}\otimes\ket{1}_{p,\boldsymbol{k}=0}$). This two-particle spinwave can be represented with a wavefunction $\Psi(x_d,x_p)$, as presented in Fig.~\ref{fig:blend}\subfig{c}, where $x_p$ is the position of an atom excited to $\ket{p}$ state, and $x_d$ is the position of an atom in $\ket{d}$ state. For simplicity, in the Figure we display the wavefunction only taking into account a single spatial dimension.

Initially, $\Psi(x_d,x_p) = \sqrt{p(x_d)p(x_p)}$, where we assume a gaussian probability $p(x)$. This depiction encodes all the possible positions that the two excited atoms could take, an example is presented in Fig.~\ref{fig:blend}\subfig{d}. Under the previously mentioned dipolar interactions, a part of the wavefunction will acquire additional phase. This is because the spinwave is symmetric; swapping the positions of individual, atomic excitations will not affect the atomic state. The region, where this phase is substantial, is depicted in purple. A slice of the wavefunction, where only the atom in state $\ket{p}$ has fixed position, is represented in Fig.~\ref{fig:blend}\subfig{e}. 

The intensity of the light recovered from atoms by reversing the LAI is given by $|\mathcal{E}_d|^2 \propto \bra\psi \hat{d}_{\boldsymbol{k}=0}^\dagger \hat{d}_{\boldsymbol{k}=0}^{ } \ket\psi$, where $\ket{\psi}$ is the state of the spinwave. If we now take into account all available spatial dimensions, we can derive (in \hyperref[sec:two_exc]{Methods} section) the formula for expected number of read-out excitations in mode $d$:
\begin{equation}
    \braket{\hat{d}^\dagger_{\boldsymbol{k}=0} \hat{d}^{}_{\boldsymbol{k}=0}}\! =\!\! \int\!\!\! p(\mathbf{x}) \Diff3\mathbf{x} \left|\int\!\!\! p(\mathbf{y}) \exp\left(-\frac{\iu t}{\hbar} V(\mathbf{x}-\mathbf{y})\right) \Diff3\mathbf{y}\right|^2
\end{equation}
Where $\mathbf{x}$ and $\mathbf{y}$ are positions of excitations in state $\ket{d}$ and $\ket{p}$ respectively. Now, the read-out gets strongly suppressed depending on this phase. The affected elements of the collective superposition will be recovered less efficiently. This means, however, that these dephased excitations will remain in the atomic cloud, contributing to the evolution dynamics, but they will not be detectable. In essence, these components will no longer fulfil the phase-matching condition and will not be efficiently read out. From now on, we will omit index $\boldsymbol{k}$, assuming that creation operators $\hat{d}^\dagger$ and $\hat{p}^\dagger$ refer only to the phase-matched spinwaves. 
From the two-particle model, we observe that there is an apparent interaction-induced decay of the spinwave excitations. The model predicts the value of this decay rate to be:
\begin{equation}
    \gamma=\frac{8 \pi^2}{9 \sqrt{3}}  \frac{C_3}{\hbar\mathcal{V}}
    \label{eq:gammath}
\end{equation}
where $\mathcal{V}$ is the effective volume of the atomic ensemble. Since the dominant type of interactions is dipolar, this additional decay is strongly suppressed if all excitations are in either $\ket{p}$ or $\ket{d}$ state.

In the experiment, we use a weak coherent state to excite on average $n_0$ Rydberg spin-wave excitations in the internal state $|d\rangle$, yielding $\ket{\psi_0} = |\sqrt{n_0},0\rangle$. After the Rabi rotation, the system achieves a bi-coherent, separable state $\ket{\sqrt{n_0} \cos(\theta/2),\iu\sqrt{n_0} \sin(\theta/2)}$, still with mean number of excitations $n_0$. 

The use of coherent states motivates us to extend the simplified two-particle model to also include multiple excitations. The details are described in the \hyperref[sec:two_exc]{Methods} section. From the multi-particle model, we observe that each excitation in the $d$ mode contributes to the dephasing of excitations in the $p$ mode, and vice versa. The excitation-number averages in both modes are:
\begin{equation}\label{eq:operator_evolution}
\begin{split}
\begin{multlined}[t][0.8\columnwidth]
\braket{n_d} = n_0 \cos^2\tfrac{\theta}{2} \exp\left[-n_0 (1-e^{-\gamma\tau})\sin^2\tfrac{\theta}{2}\right]\\
\approx n_0 \cos^2\tfrac{\theta}{2} \exp\left(-n_0 \gamma\tau\sin^2\tfrac{\theta}{2}\right)
\end{multlined}\\
\begin{multlined}[t][0.8\columnwidth]
\braket{n_p} = n_0 \sin^2\tfrac{\theta}{2} \exp\left[-n_0 (1-e^{-\gamma\tau})\cos^2\tfrac{\theta}{2}\right]\\
\approx n_0 \sin^2\tfrac{\theta}{2} \exp\left(-n_0 \gamma\tau\cos^2\tfrac{\theta}{2}\right)
\end{multlined}
\end{split}
\end{equation}
Where $\tau$ is the interaction time. Here we expect $\gamma \tau \ll 1$ (in the experiment $\gamma \tau \approx \num{0.04}$), which yields an approximate version with the simple exponential decay from Eq.~\ref{eq:operator_evolution}. The magnitude of the interactions depends on the time $\tau$ the two Rydberg excitations interact with each other, which is limited by the memory lifetime. The larger the interaction time, the stronger the effect of interaction on the stored spinwaves.
As a result, the Rabi oscillation of the population between the two collective Rydberg states becomes much steeper; we will refer to them as super Rabi oscillations.

At the end of the experiment, we use LAI to convert the atomic excitations back into photons and measure them with single-photon detectors. The conversion happens with efficiency $\eta$.
\begin{figure*}[t]
\centering
\includegraphics[width = 2\columnwidth]{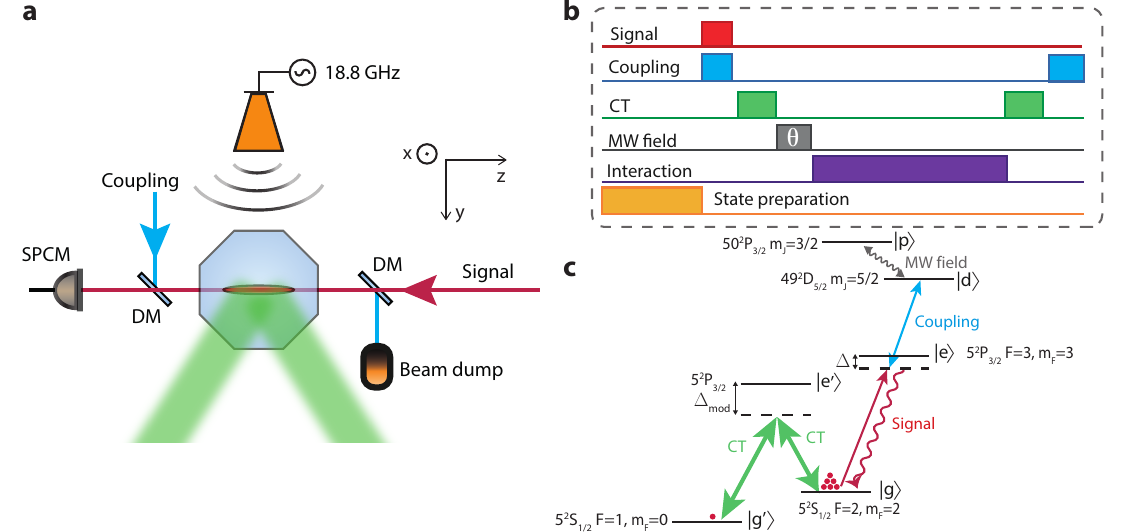}
\caption{\label{fig:mot_lvls_seq}
\subfig{a} Scheme of the experimental setup based on the ultracold atoms in the magneto-optical trap and geometric arrangement of the beams. The signal beam and counter-propagating coupling beam are exciting the spinwave in the atomic ensemble. Coherence transfer (CT) beams shining from the side of the chamber are cancelling the wavevector of the induced coherence. A microwave field is applied from the side with a horn antenna. 
\subfig{b} Experimental sequence for the interaction-enhanced metrology via increasing the steepness of the Rabi oscillation slope. 
\subfig{c} Relevant $^{87}\mathrm{Rb}$ energy level configuration for the experiment.
}
\end{figure*}

\begin{figure*}[t]
\centering
\includegraphics[width = 2\columnwidth]{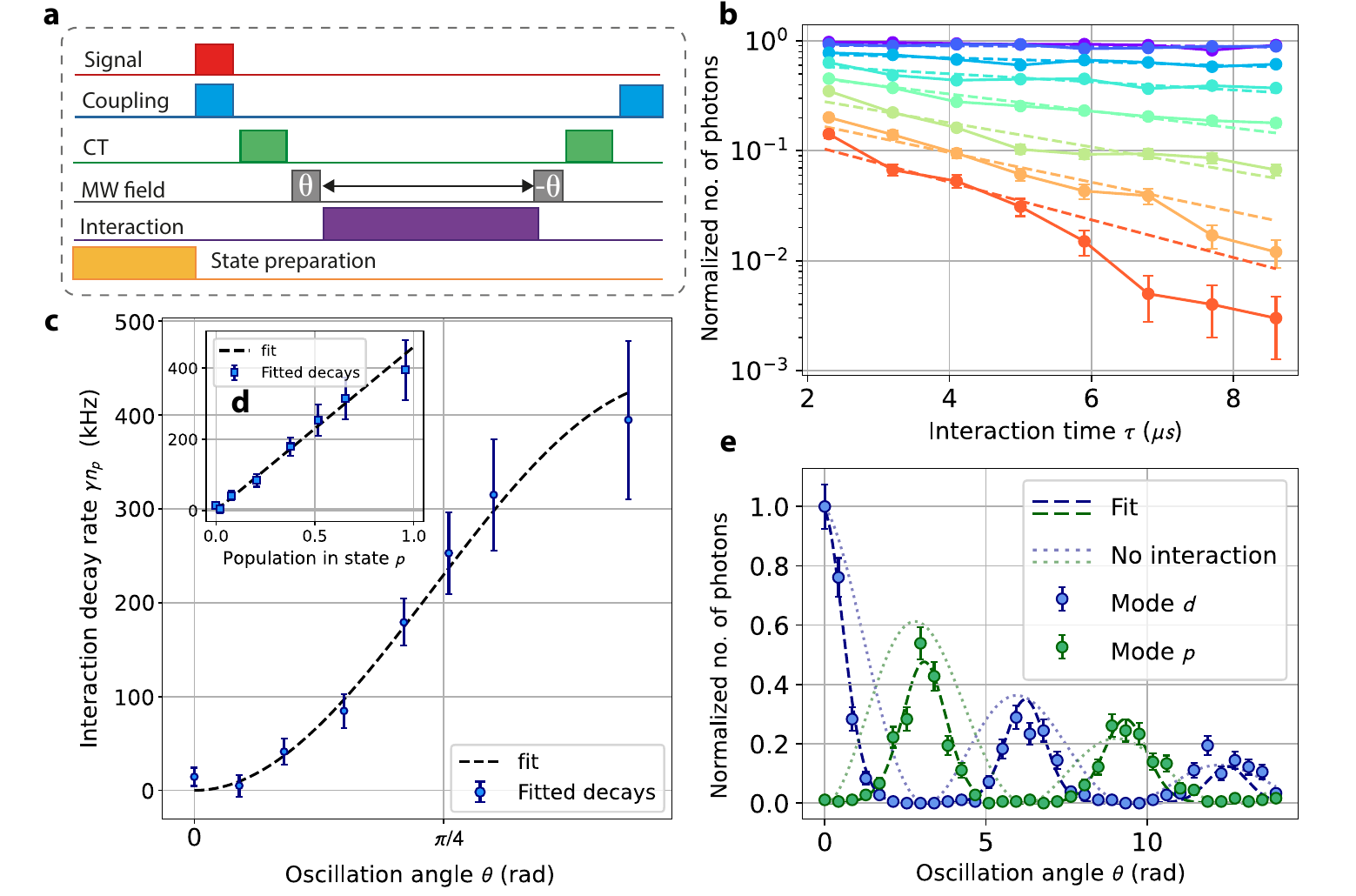}
\caption{\label{fig:exp_data_all}
\subfig{a} Experimental sequence for the measurement of the interaction-induced decay rate as a function of the super Rabi oscillation angle $\theta$.
\subfig{b} Data for calibration of interaction-induced decay rate with fitted exponential decays for different super Rabi oscillation angles.
\subfig{c} Interaction-induced decay rates with fitted theoretical predictions.
Inset of \subfig{c}: \subfig{d} Interaction-induced decay rates as a function of the ratio of population in the $p$ mode with fitted linear function.
\subfig{e} Rabi oscillation between mode $d$ and $p$ modified by interaction-induced decay. Blue points represent oscillations measured at mode $d$. Green points represent oscillations measured at mode $p$. Dashed lines are theoretical predictions that were fitted to experimental data. Dotted lines represent Rabi oscillation without Rydberg dipolar interactions. The experimental data were normalised with the normalisation factor $\bar{n}$. 
Plots \subfig{b} and \subfig{c} correspond to the experimental sequence depicted in \subfig{a}, and plot \subfig{e} corresponds to the experimental sequence depicted in Fig. \ref{fig:mot_lvls_seq}\subfig{b}.
 }
\end{figure*}

\begin{figure*}[t]
\centering
\includegraphics[width = 2\columnwidth]{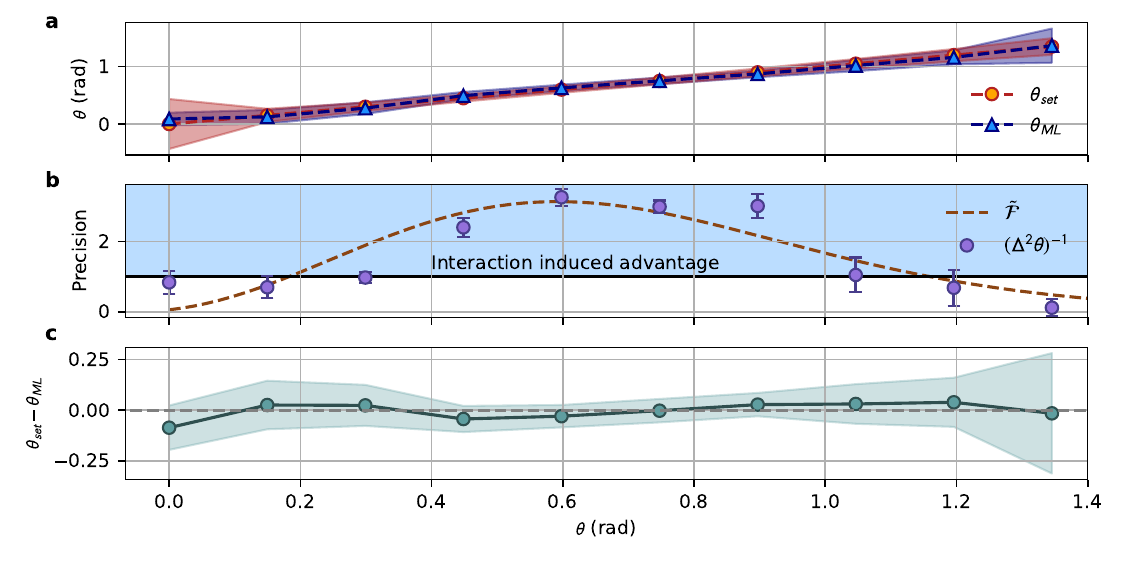}
\caption{\label{fig:ML}\subfig{a} Estimated oscillation angle calculated using maximum likelihood estimation is represented with blue triangles. Set oscillation angles are represented with orange dots. Blue and red shaded areas represent the variance of the maximum likelihood estimator and inverse square root of FI, respectively, per $N = 100$ experimental shots. \subfig{b} Normalized theoretical FI and inverse of normalized variance of the estimator. The black line represents the value of the normalized FI without the interactions, and the shaded area above represents the interaction-induced metrological improvement. \subfig{c} Difference between the set oscillation angle and the estimated value.}
\end{figure*}

\subsection{Experimental setup}
The quantum memory setup is based on the cold $^{87}\text{Rb}$ ensemble generated in the magneto-optical trap (MOT) as depicted in Fig.~\ref{fig:mot_lvls_seq}\subfig{a}. The atoms form a $\SI{1}{\cm}$-long, cigarette-shaped cloud with optical depth OD = 190. The experiments are performed in sequences lasting \SI{12}{\ms} each, synchronized with mains frequency. 
The atoms are prepared in the state $\ket{g}$ and the initial coherence is induced using two-photon absorption of the signal enabled by the coupling beam tuned to the two-photon resonance with a Rydberg state $\ket{d}$. Configuration of energy levels used in the experiment is depicted in Fig.~\ref{fig:mot_lvls_seq}\subfig{c}. Large mismatch of the signal and coupling wavevectors reduces the lifetime of the coherence to $\tau_{\text{unmod}} = \SI{2.5}{\micro \s}$ due to residual thermal motion. To overcome this effect, we transferred the ground state part of the coherence to the state $\ket{g'}$ with the use of the crossed coherence transfer (CT) beams, improving the lifetime to \SI{31}{\micro \s}.   
Explanation of the CT method, along with a detailed description of the experimental setup, is available in the \hyperref[sec:experiment]{Methods}.

%%%
To experimentally verify the theoretical model of the interacting spinwaves, we measured the value of interaction-induced decay rates of the excitations in mode $d$ as a function of the oscillation angle $\theta$. The protocol for this measurement is presented in Fig.~\ref{fig:exp_data_all}\subfig{a}. The storage time of the Rydberg spinwave was set to a constant value $t_S =\SI{10}{\micro \s}$. To invoke the dipolar interactions, the atoms were placed in a superposition between two states with the MW field pulse performing Rabi rotation by the angle $\theta$. We applied the second MW pulse with the opposite phase right before the read-out. To measure the interaction-induced decays at different interaction times $\tau$, the MW pulse separation was varied. The interaction-induced decay rates for the different oscillation angles are depicted in Fig.~\ref{fig:exp_data_all}\subfig{c}. The exponential decays as a function of interaction time with fitted theoretical predictions according to Eq.~\ref{eq:operator_evolution} are presented in Fig.~\ref{fig:exp_data_all}\subfig{b}, different colors correspond to different oscillation angles $\theta$. 
In Fig~\ref{fig:exp_data_all}\subfig{d} we have presented the interaction-induced decay rates as a function of the initial number of excitations in mode $d$. As expected, they are proportional to the number of excitations in mode $p$. The measured decay rate calculated per a single sensor particle in state $\ket{p}$ is $\gamma = \SI{1.4(0.3)}{\kHz}$, which agrees closely with the theoretically predicted value $\gamma_\text{th} = \SI{1.9(0.5)}{\kHz}$.

Within the introduced framework, we now realize the protocol for measuring the super Rabi oscillations. The number of signal photons stored in the memory is set such that the mean number of read-out photons for $\theta = 0$ and without interactions is $\eta n_0=\bar{n}=\num{1.1(0.1)}$ per sequence to ensure only long-range interactions and neglect the van der Waals type interactions.
The interaction time is set to $\tau = \SI{20}{\micro\s}$ such that it is much longer than the duration of the MW pulse. The total efficiency of the photon detection was determined to be $\eta_{det} = 2\%$.
The super Rabi oscillation angle is changed by changing the duration of the MW pulse with constant Rabi frequency $\Omega_{MW} = 2\pi \times \SI{0.66}{\mega \hertz}$ according to the sequence presented in Fig.~\ref{fig:mot_lvls_seq}\subfig{b}. In Fig.~\ref{fig:exp_data_all}\subfig{e}, we show the super Rabi oscillation with the fitted theoretical predictions of the mean number of detected photons corresponding to the number of excitations in modes $d$ and $p$. To sample the number of excitations in $p$ mode, we added an MW $\pi$-pulse before the read-out. The dotted lines act as a reference of the standard Rabi oscillations without the Rydberg interactions. Note that even excluding the fitted additional exponential decay, super oscillations at both modes do not sum up to unity for every angle, contrary to the standard oscillations.

\subsection{Field estimation via Rabi oscillations}

To benchmark the metrological advantage of the presented protocol, we utilised the FI approach. In this way, we can bound the limit of the estimation precision of the oscillation angle, which corresponds to the limit of the weakest possible detectable MW field. This is captured by the Cramér-Rao bound: 
\begin{equation}\label{eq:fisher_information}
    \Delta^2\theta \geq \frac{1}{\mathcal{F}_{\theta}} , \
    \mathcal{F}_{\theta} = \sum_n \frac{1}{P_{\theta}(n)}\left(\frac{\partial P_{\theta}(n)}{\partial \theta}\right)^2  
\end{equation}
The analytical formula for the FI for the super Rabi oscillation can be derived from the multi-particle model of interacting spinwaves by calculating the probability distribution $P_\theta(n)$ of the number of excitations in mode $d$ (see \hyperref[sec:phenomenological]{Methods}). From the multi-particle model, we also note that with sufficient values of $\gamma \tau$, it is enough to measure the number of excitations in one of the modes for estimation. 
To obtain the greatest advantage over the detection losses, the choice of $\tau$ has to be carefully balanced between estimation gain and finite lifetime.
The precision of the experimental measurements can be described with maximum likelihood (ML) estimation, which allows for the quantification degree to which the Cramér-Rao bound is saturated by the proposed scheme.
We define the ML estimator as:
\begin{equation}
    \tilde{\theta}_{\text{ML}}(n) = \underset{\theta}{\text{argmax}}[l_n(\theta)]
\end{equation}
where $l_n(\theta) = P_{\theta}(n)$ is the likelihood function.
To calculate the ML estimator of the oscillation angle, for each set angle $\theta$, we capture $N_0 = 10000$ experimental shots. We divide the measurements evenly into $k$ realizations with $N$ shots each, such that $N\times k = N_0$. For each realization, we estimate the value of $\theta_\text{ML}$, from which we get statistics for $k$ realizations and calculate the variance $\Delta^2 \theta$. The FI per shot is calculated from the variance as $\mathcal{F} = 1/(N\Delta^2\theta)$. To recover the error of ML estimator variance (and FI), we bootstrap the data by changing the division method of $N_0$ between $N$ and $k$.

Enhancement in FI can be quantified by introducing a normalized FI, which compares the FI measured in the system and the value of FI without interactions $\tilde{\mathcal{F}} = \mathcal{F}/n_0\eta = \mathcal{F}/\bar{n}$.
In Fig.~\ref{fig:ML}\subfig{a}, we compare the mean value of the ML estimator with the value of the set oscillation angle. The red shaded area represents the inverse of the FI for the experimental parameters obtained from the analytical formula. The blue shaded area corresponds to the variance of the ML estimator. For clarity, both values are shown per $N=100$ shots.
Fig.~\ref{fig:ML}\subfig{c} shows the bias of the estimator, the shaded area is the variance of the ML estimator.

Fig.~\ref{fig:ML}\subfig{b} compares the theoretical value of normalized FI with the normalized inverse of the variance of the ML estimator $\Delta^2 \theta$. The black line represents the standard value of the normalized classical FI, that is, without the interaction-induced enhancement. 
The analytical formula for the FI shows the improvement of the amount of information carried by the photon in the lossy detection. Here, it can be seen that the normalized precision of the ML estimation obtained from the experimental values follows the theoretical predictions very well.
We achieved FI of $\mathcal{F} = \num{3.6(0.3)}$, corresponding to a normalized FI of $\tilde{\mathcal{F}} = \num{3.3(0.3)}$, over a threefold improvement in measurement precision compared to the standard limit of FI for lossy measurement (note that we do not exceed the value of FI without losses).

The improvement in the estimation precision allows for benchmarking the detection capabilities of the presented protocol. 
We calculate the minimal variance of the MW Rabi frequency (per experimental shot), which translates to the minimal variance of the detectable electric field, according to the formula 
\begin{equation}
    \sqrt{\Delta^2 E_{MW}} = \Delta E_{MW} = \frac{\sqrt{\Delta^2 \theta}}{T}\frac{\hbar}{d_{dp}} 
\end{equation}
where $T$ is the MW pulse duration corresponding to the oscillation angle with the highest estimation precision and $d_{dp}$ is the transition dipole moment between $\ket{d}$ and $\ket{p}$ states. We were able to achieve the estimation precision $\Delta E_{MW} = \SI{44}{\micro \V \cm ^{-1}}$ per single experimental shot.

Rydberg sensors operating in a continuous mode are characterised by the noise spectral density rather than the precision achieved per a single shot. 
If we treat the atomic ensemble as the MW field detector, we can also define its best sensitivity $S_{E_{MW}} = \Delta E_{MW}\sqrt{T} = \SI{39}{\nano \V \per \cm \hertz\tothe{-1/2}}$ assuming a 100\% duty cycle of signal acquisition. A high duty cycle of at least 50\% would be practically feasible.

\subsection{Toy Model}
\begin{figure*}
    \centering
    \includegraphics[width=\textwidth]{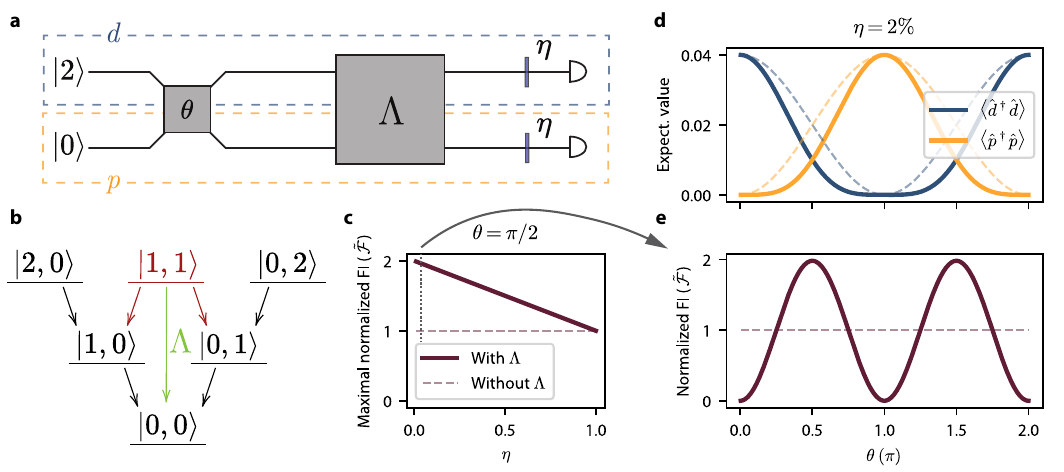}
    \caption{
    \subfig{a} Scheme of the precision enhancement protocol. After encoding the parameter $\theta$, the system undergoes a pre-measurement processing operation denoted with $\Lambda$. Later, both modes $p$ and $d$ are measured with efficiency $\eta$.
    \subfig{b} Action of a lossy channel on the states. The proposed interaction $\Lambda$ transfers the red component $\ket{1,1}$, which introduces overlapping decay paths to state $\ket{0,0}$ and leaves black components with reversible decay paths.
    \subfig{c} Increase of FI with the proposed interaction for a range of detection efficiencies $\eta$. The dotted line corresponds to the value of $\eta$ presented in \subfig{e}.
    \subfig{d} The plot shows expectation values of excitation number operators on the state. The dashed lines represent expectation values without the operation.
    \subfig{e} The plot shows the normalized FI and its enhancement for some values of the encoded parameter. The dashed line represents FI without the operation. Normalized FI compares the performance with and without the operation $\Lambda$.
    }
    \label{fig:sim}
\end{figure*}
For the case of two excitations, we developed an intuitive toy model that demonstrates a similar advantage in precision of estimation via a simple operation. The multi-particle model that we used to describe the experimental setting converges to the presented toy model in the limit of large $\gamma \tau$. More details are presented in section \hyperref[sec:conn]{Methods}.

We assume that we are working in a Hilbert space limited to at most two bosonic excitations distributed among two modes with corresponding creation operators $\hat{d}^\dagger$ and $\hat{p}^\dagger$.
We use spinwaves -- collective Rydberg excitations, where one excitation can be represented as a collective superposition $\ket{1,0} = \hat{d}^\dagger \ket{0,0} = 1/\sqrt{N} \sum_k \ket{g_1\ldots d_k \ldots g_N}$ of $k$-th atom (out of $N$ atoms) being in $\ket{d}$ state and the rest in ground state $\ket{g}$.

The protocol is represented by a diagram in Fig.\ref{fig:sim}\subfig{a}. 
The system is initially prepared in state $\ket{\psi_0} = \ket{2,0} = \hat{d}^{\dagger}\hat{d}^{\dagger}/\sqrt{2} \ket{0,0}$. The estimated parameter $\theta$ is encoded in a unitary operation taking the form $U(\theta) = \exp\left(- \iu\, \theta/2 \sum_k \hat{\sigma}_{dp}^{k} + h.c  \right)$, where $\hat{\sigma}_{dp}^{k} = \ket{g \ldots d_k \ldots g}\bra{g \ldots p_k \ldots g}$ is a ladder operator acting on the $k$-th atom. This operation represents a Rabi rotation between internal states $\ket{p}$ and $\ket{d}$, which we associate with collective modes $p$ and $d$. For the collective state, the parameter is effectively encoded in a Bogoliubov transformation of the operators: $\hat{p}\rightarrow\cos(\theta/2)\hat{p}+i\sin(\theta/2)\hat{d}$ and $\hat{d}\rightarrow\cos(\theta/2)\hat{d}-i\sin(\theta/2)\hat{p}$. For our initial $|2,0\rangle$ state, the parameter encoding in general results in a superposition of all states from the two-particle manifold:
\begin{equation}
    |\psi_\theta\rangle=\cos^2\left(\frac{\theta}{2}\right)|2,0\rangle + \frac{i}{\sqrt{2}}\sin(\theta)|1,1\rangle - \sin^2\left(\frac{\theta}{2}\right)|0,2\rangle
    \label{rotated20state}
\end{equation}
Information about the parameter is now encoded in the amplitudes of the above state, and simple measurement of populations $\hat{p}^\dagger \hat{p}$ and $\hat{d}^\dagger \hat{d}$ is then enough to optimally estimate $\theta$ and saturate the quantum Cramer-Rao bound \eqref{eq:qcrb}. The FI in such measurement equals the QFI for this state $\mathcal{F}=\mathcal{F}_Q=2$, meaning that the choice of measurement base is optimal.

Let us now include detection losses in the model  
and analyse their effect on various terms in Eq.~\eqref{rotated20state} (see Fig.~\ref{fig:sim}\subfig{b}). 
In particular, provided that at most a single particle is lost, the $\ket{2,0}$ and $\ket{0,2}$ components 
follow distinguishable decay paths, and hence if only these components were present, information encoded in amplitudes of these terms would not be lost. At the detection stage, this would lead to a lower number of photons detected but no loss of sensitivity.  However, if  $\ket{1,1}$ component is present, its single-particle decay process leads to the same output states as in case of the remaining components, making it impossible to recover full information on the original amplitudes and resulting in $\mathcal{F} = \eta \mathcal{F}_Q = 2\eta$ characteristic for reduction of FI in lossy detection and indepdent probes/excitations, see \hyperref[sec:methods]{Methods}.

From this perspective, it is possible, that one may be better off preemptively removing the $\ket{1,1}$ component (even if there was some non-zero information on the parameter in its amplitude), and in this way can recover full information encoded in the amplitudes of $\ket{0,2}$ and $\ket{2,0}$ components, despite presence of loss.

Following this intuition, let us consider a non-unitary operation $\Lambda$, involving interaction (non-linear losses) between excitations in different modes, that is parametrized by Kraus operators 
\begin{equation}
\begin{aligned}  
\hat{K}_0 &= \ketbra{0,0}{1,1} \\  
\hat{K}_1 &= \identity - \ketbra{1,1}{1,1} 
\end{aligned}
\end{equation}
and is applied after the parameter encoding stage. 
 The result of the interaction is depicted as a green arrow in Fig.~\ref{fig:sim}\subfig{b}. The proposed operation transfers the troublesome component $\ket{1,1}$ to $\ket{0,0}$,
 while preserving components with distinct decay paths ($\ket{2,0}$ and 
 $\ket{0,2}$). For $\eta=2\%$ this protocols results in a steepened expectation values of excitation numbers as shown in Fig.~\ref{fig:sim}\subfig{d}. The output state exhibits a remarkable increase in the precision of estimation of $\theta$, with the optimal performance around $\theta=\pi/2$ (visible in Fig.~\ref{fig:sim}\subfig{e})
 for which $\mathcal{F} = 2 \eta(2-\eta) \geq 2\eta$, as depicted in Fig.~\ref{fig:sim}\subfig{c}, see \hyperref[sec:methods]{Methods} for the derivation of the formula. Note that this strategy only modifies the scaling of FI with losses; the resulting FI does not exceed the initial value. We then compared the obtained scaling with the one numerically found by an optimization procedure implemented in QMetro++ \cite{pdulian2025pdulian} package (the script has beed deposited along with the data). Remarkably, with the help of the fundamental bound from \cite{PRXQuantum.4.040305}, we prove that the proposed strategy is indeed optimal for the model considered, see \hyperref[sec:methods]{Methods}.

We should stress that, despite some apparent resemblance to post-selection metrology protocols \cite{ArvidssonShukur2020}, where 
FI of a post-measurement state may be conditionally enhanced depending on the results of some 
filtering measurement; here we \emph{do not} post-select any events, and honestly compute the full FI including all the events that may be observed. It should be, therefore, appreciated even more that despite this `more demanding' paradigm, we still observe an enhancement.

\section{Discussion}
In this work, we have demonstrated the interaction-enhanced metrology scheme, which allows for error prevention in experiments with lossy measurement schemes.
We have theoretically shown the advantage of introducing the additional non-linear losses before the detection losses for the two-particle model. We quantified the advantage of this operation using the QFI approach and showed that it is the optimal operation for the two-excitation system.

To experimentally realize the error prevention operation, we have presented the model of two interacting dipoles, allowing us to quantify the intrinsic interaction-induced decay between the pair of Rydberg excitations. The theoretically predicted decay rate $\gamma$ is consistent with the experimentally measured values, allowing us to propose the multi-particle model to quantify the evolution of coherent state stored in the quantum memory (see \hyperref[sec:methods]{Methods} for details), allowing for the interaction-enhanced metrology protocol.
The short lifetime of the Rydberg spinwaves is extended via momentum-cancelling coherence transfer achieved with two crossed state-selective beams. 
The resulting long coherence time allowed the interactions between Rydberg pairs to resurface, and the phenomenon of super Rabi oscillations was observed.
The theoretically predicted advantage, manifested as a steeper decay of oscillations, allows estimation of the precision of measurement. Using maximum likelihood estimation, we compared the estimated precision with the theoretical limit introduced by the Cramér-Rao bound.
We have achieved the FI per detected photon $\mathcal{F} = \num{3.3(0.3)}$, showing over threefold improvement as compared to the standard procedure without additional operation before detection.

While current parameters allow for demonstrating the efficacy of the protocol, numerous aspects could be explored more to increase the advantage and improve the efficiency of the experiment.
The first obvious improvement could be increasing the optical depth of the atomic ensemble to increase the efficiency of the protocol. This would also allow for the generation of more Rydberg spin waves, allowing for a larger improvement in the FI per photon.
Also, the temperature of the atomic memory could be lowered to reduce the residual thermal motion. 
At the end, fully harnessing the potential of Rydberg lifetime extension protocol would require a cryogenic environment \cite{CantatMoltrecht2020,Muni2022}. 

While the presented sensitivity of \SI{39}{\nV\per\cm\Hz\tothe{-1/2}} is on par with the best Rydberg sensors based on hot atoms, it could be straightforwardly improved by using more atoms and exciting more spin waves in the ensemble, which requires better ensemble preparation and more power of the control beam than has been available in this study. 
Apart from raw performance improvements, we envisage two major development directions for our research. 
Firstly, more advanced sensing protocols and secondly, enhancing the role of interactions up to sensing singular excitations.
For a sufficiently strong decay rate $\gamma$, we will be able to distinguish cases of one versus zero sensor particles being excited (Rabi-rotated) by the external microwave field. This presents an overarching opportunity to design quantum-enhanced microwave sensors operating in the fully quantum regime. 

\section{Data Availability}
Data that supports this study and a script used for numerical optimisation of the operation in the toy model have been deposited at \cite{MJMXI8_2025} (University of Warsaw research data repository).

\section{Acknowledgments}
\begin{acknowledgments}
This research was funded in whole or in part by the National Science Centre,
Poland grant No. 2021/43/D/ST2/03114.
 The "Quantum Optical Technologies" (FENG.02.01-IP.05-0017/23) project is carried out within the Measure 2.1 International Research Agendas programme of the Foundation for Polish Science, co-financed by the European Union under the European Funds for Smart Economy 2021-2027 (FENG). We thank J. Kołodyński and M. Papaj for insightful discussions.
 We also thank P. Dulian for discussions about the QMetro++ code \cite{pdulian2025pdulian}.
\end{acknowledgments}

\section{Author Contributions}
SK and BN conducted the experiment and analysed the data, assisted by MM, WW and MP. All authors contributed to the development of various parts of the theory and the writing of the paper. MM, WW and MP conceived and led various parts of the project, while RDD identified the ultimate sensitivity bounds. MP coordinated the project.

\bibliographystyle{apsrev4-2}
\bibliography{refs}

\section{Methods}
\label{sec:methods}

\subsection{Experimental setup} \label{sec:experiment}
\paragraph{Preparation}
To conduct the experiment, we used a cloud of $N\approx10^8$ ultra-cold $^{87}$Rb atoms acting as an atomic memory.
Atoms form an elongated cigar-shaped cloud measuring $\SI{0.4}{\mm} \times \SI{0.4}{\mm} \times \SI{9}{\mm}$ with optical depth reaching 190 on the relevant atomic transition $\ket{g} \rightarrow \ket{e}$, and atomic density $n\approx\SI{10e10}{\cm^{-3}}$.
The atomic ensemble is cooled to a temperature of \SI{78}{\micro\K} and captured using a magneto-optical trap (MOT).
The experiments are performed in sequences lasting \SI{12}{\ms} each, synchronized with mains frequency. 
To energetically separate relevant levels for the protocol, for the duration of the experiment, the trap is held in the bias magnetic field $\boldsymbol{B} = B_0\hat{z}$, where $B_0 \approx \SI{0.85}{G}$. MOT coils and cooling beams are turned off for the experiment. 
After cooling and trapping procedures, the atoms are optically pumped to the state $\ket{g} = \ket{5^2S_{1/2}\, F = 2, m_F = 2}$. 
This is done by illuminating the atoms for $\SI{12}{\us}$ with two laser beams. The first beam is $\pi$-polarized and utilizes $\ket{5^2S_{1/2}\, F = 1} \rightarrow \ket{5^2P_{1/2}\, F = 2}$ transition, effectively emptying the $\ket{5^2S_{1/2}\, F = 1}$ manifold. The second laser propagates along the atomic ensemble (axis of quantization) with $\sigma_+$ polarization and is tuned to the $\ket{5^2S_{1/2}\, F = 2} \rightarrow \ket{5^2P_{3/2} F = 3}$ transition, ideally pumping all atoms to the $\ket{g}$ state.

We store the signal in the Rydberg atomic coherence by shining the atoms with the signal laser red detuned with the $\ket{g}$ to $\ket{e} = \ket{5^2P_{3/2}\,F = 3, m_F = 3}$ transition by $\Delta = \SI{30}{\mega \hertz}$ and with the coupling laser tuned to two-photon resonance with the Rydberg state $\ket{d} = \ket{49^2D_{5/2}\,m_J = 5/2}$. In this way, we have created ground-Rydberg atomic coherence which is extremely sensitive to the MW field because of the large dipole moment $d_{dp} = 1950 e a_0 $. The blockade radius for $\ket{d}$ is $r_b=\SI{2.1}{\micro \m}$. We set the waists of the coupling and signal beams in the cloud's near field to be respectively \SI{80}{\um} and \SI{70}{\um}. 

\paragraph{Coupling beam generation}
The coupling beam was generated with $\SI{960}{\nano \meter}$ tapered amplifier (Moglabs Optical Amplifier), seeded by an External-Cavity Diode Laser (ECDL, Moglabs cateye), and frequency doubled with 50~mm MgO:LiNbO$_3$ PPLN crystal in the single pass configuration. The seed laser was locked to the independent \SI{1560}{\nano \meter} narrowband fiber laser via cavity transfer lock. 
 
\begin{figure}[t]
\centering
\includegraphics[width = 1\columnwidth]{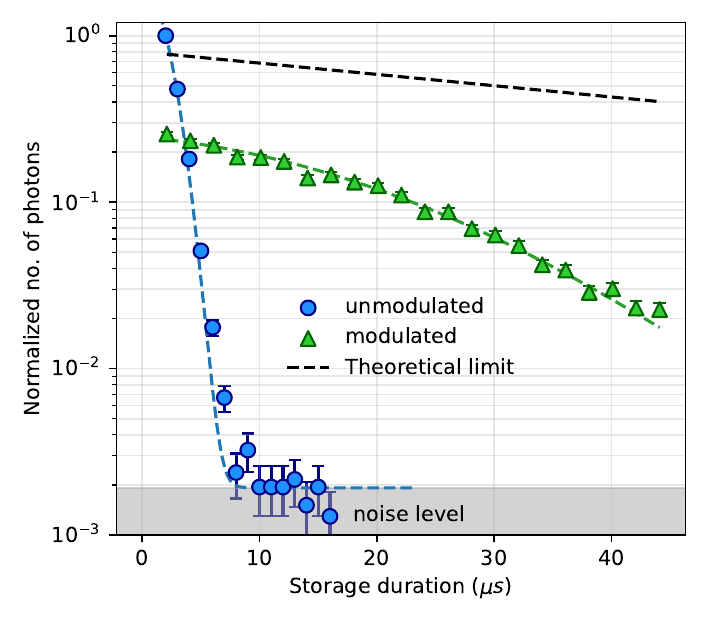}
\caption{\label{fig:lifetime}(Extended Data Figure) Rydberg memory lifetime for standard and modulated protocol. Blue dots represent the normalised intensity of unmodulated stored signal as a function of storage time. Green triangles represent the normalised intensity of the stored signal modulated by shortening its spin-wave wavevector. The dashed grey line is the maximal possible theoretical limit of the modulation efficiency that takes into account imperfect polarizations due to the rotation of beams.}
\end{figure}

\paragraph{Momentum cancelation}
Atomic coherence induced by signal field ($\SI{780}{\nano \meter}$) and control field ($\SI{480}{\nano \meter}$) has wavevector equal to the difference between wavevectors of the incident fields $|\boldsymbol{k_s}| = |\boldsymbol{k_p} - \boldsymbol{k_c}| =2\pi \times\SI{0.8}{\micro \meter ^{-1}}$. 
Due to the large wavevector mismatch between signal and coupling beams, the thermal lifetime of the produced spinwave is very short $\tau_{\text{unmod}} \approx \SI{2.5(0.1)}{\micro \s} $. To mitigate this problem, we change the wavevector of the created atomic coherence by removing its momentum with two coherence transfer (CT) beams. 
This is done by shining the atomic ensemble with the two crossed beams immediately after the spinwave is created. First beam is red detuned by $\Delta_\text{mod} = \SI{5}{\giga \hertz}$ from the $\ket{g} \rightarrow  \ket{e'} = \ket{5^2P_{3/2}\,F = 3}$ transition. The second beam is tuned to the two-photon transition between $\ket{e'}$ and $\ket{g'} =  \ket{5^2S_{1/2}\, F = 1, m_F = 0}$. The angle between two beams has been set to a magic angle $\alpha = \ang{37}$, such that the resultant wavevector of the CT beams is equal to the wavevector of the stored Rydberg spinwave. Before read-out, this procedure has to be reversed. This way, we can reach a lifetime greater by an order of magnitude $\tau_{\text{mod}} = \SI{31(1)}{\micro \s} $. The comparison between modulated and unmodulated memory lifetime is depicted in Fig.~\ref{fig:lifetime}. The CT efficiency was measured to be $\eta_{mod} = 26\%$.  This method has a significant advantage (over an order of magnitude) over the previously applied operation utilising the ac-Stark effect of two interfering beams \cite{Kurzyna2024longlivedcollective}, where the efficiency of the protocol was around 1\%.

Later, after the CT procedure, a MW field with linear polarization along the $x$ axis is applied from the sides of the atomic ensemble and drives the transition $\ket{d} \rightarrow \ket{p} = \ket{50^2P_{3/2}\,m_J = 3/2}$, putting the atoms in a superposition between the two Rydberg states.
The MW field is generated with the  LMX2595EVM synthesiser and has a set frequency of $f=\SI{18.8}{\GHz}$. After the MW pulse, we allow the collective atomic state to evolve under dipolar Rydberg interactions for a variable time. Next, we reverse the CT procedure and apply the coupling laser once again to stimulate the read-out of the collective excitations. Later we send the read-out through an optical setup and a bandpass filter and then coupled it to a fiber detector. The read-out has been measured at a single-photon level using a single-photon counting module (SPCM-AQRH-14 Excelitas) with a quantum efficiency of 60\%.

\begin{figure}[t]
\centering
 \includegraphics[width = 1\columnwidth]{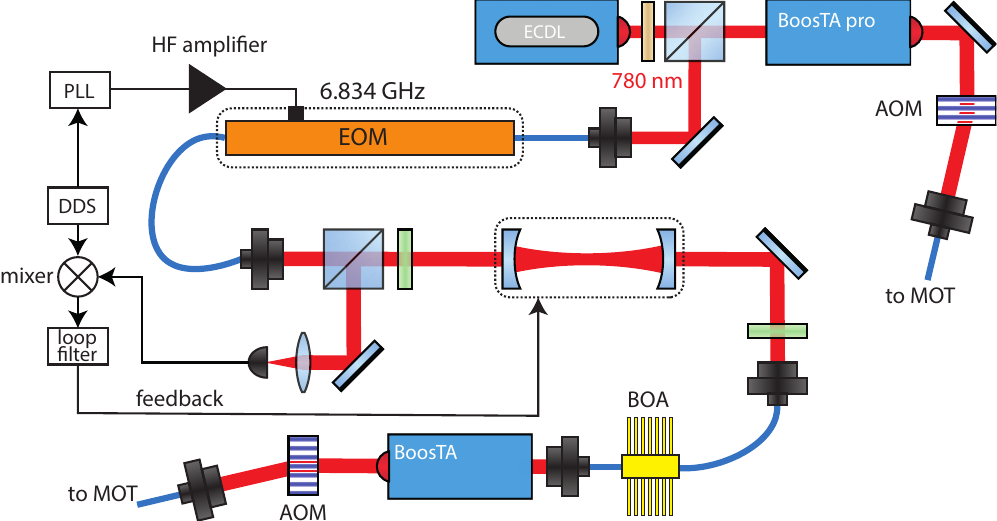}
\caption{\label{fig:sideband}(Extended Data Figure) Cavity sideband locking and amplification of the beams for the CT stage of the experiment. The two CT beams are generated from one ECDL laser. One beam is amplified with the BoosTA pro amplifier. The second beam is modulated with the electro-optic modulator (EOM) fed with an amplified $\SI{6.83}{\giga \hertz}$ RF signal. Afterwards, the sideband with the higher frequency is filtered with a Fabry-Perot cavity and sent to seed a booster optical amplifier (BOA). Finally, the second stage of amplification uses the BoosTA amplifier and its output is sent to the atomic ensemble.}
\end{figure}

\paragraph{Coherence transfer beams stabilization}
Scheme of the setup for CT beams is depicted in the Fig ~\ref{fig:sideband}.
CT beams are carved from one continuous-wave ECDL laser, frequency-stabilised by a phase-locked loop to a narrowband frequency-doubled fibre laser (NKT Photonics, $\SI{1560}{\nano\m}$). The laser beam is divided into two branches. The first one is sent directly to a tapered optical amplifier (Toptica BoosTA pro).
Pulses for this branch are generated with an acousto-optic modulator (AOM) in a double pass configuration fed with a signal with frequency $\SI{110}{\mega \hertz}$ controlled by the main field programmable gate array (FPGA) system.

The second one is frequency shifted by $\SI{6834}{\mega \hertz}$ with fiber electro-optic modulator (EOM) and filtered using a Fabry-P\'erot cavity. The filtered signal is then amplified in two stages. Firstly, by a fibre Booster Optical Amplifier (Thorlabs BOA785S) and then with a tapered optical amplifier (Toptica BoosTA). The pulses are generated with an AOM in a single pass configuration fed with the same signal from the FPGA.

\paragraph{Limits on number of Rydberg excitations}
Lifetime highly depends on the number of created Rydberg atoms; the higher the number of incident photons, the lower the lifetime due to the $R^{-6}$ interactions between Rydberg atoms. In the metrology-type experiment, we want to avoid such interactions; thus, we chose the number of excitations in such a way that, apart from spontaneous emission, the spinwave decay is strictly Gaussian and comes from residual thermal motion of the atomic ensemble.      
With a higher power of signal, the number of Rydberg atoms increases and thus the $R^{-6}$ interactions induce the undesirable exponential decay, reducing the memory lifetime. We also work in a regime where excitations are not affected by the dipole blockade.

\begin{figure}[t]
\centering
\includegraphics[width = 1\columnwidth]{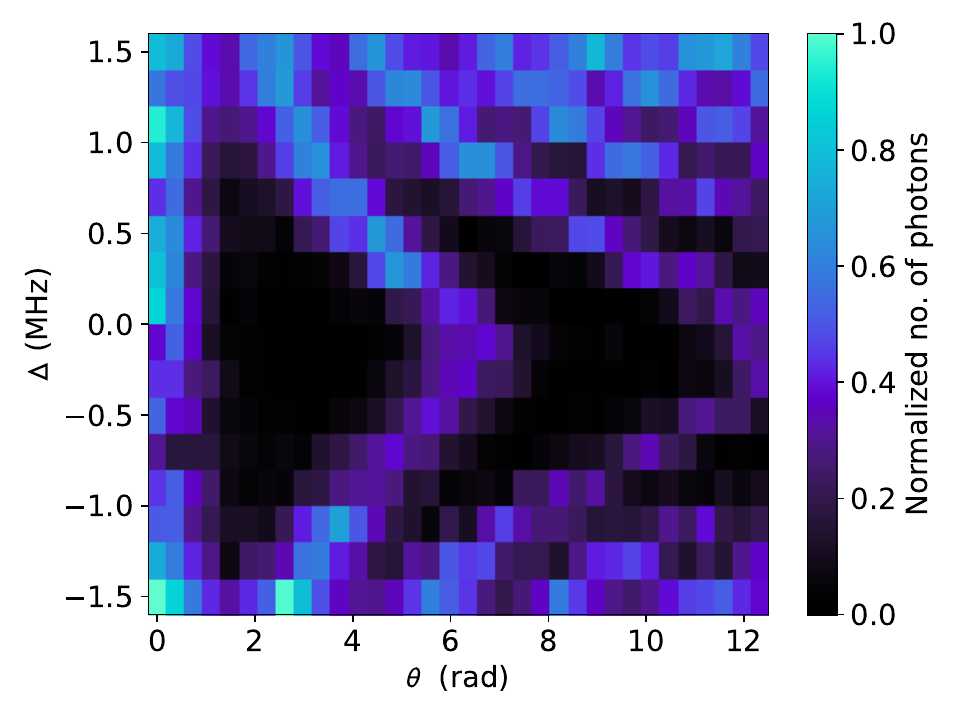}
\caption{\label{fig:rabi_map}(Extended Data Figure) Map of interaction-induced super Rabi oscillation between $\ket{d}$ and $\ket{p}$ for different detuning from the Rydberg transition resonance and super Rabi oscillation angle.}
\end{figure}

\paragraph{Off-resonant super Rabi oscillations}
We use a typical qubit characterization protocol to make sure that the microwave field is tuned on resonance. We measure Rabi oscillations for a range of pulse lengths (quantified by oscillation angle $\theta$) and detuning $\Delta$ of the microwave field from resonance. The dipolar interactions between different Rydberg states introduce a super-oscillation character to the picture. The oscillations in a function of MW field detuning from the transition resonance are shown in Fig.~\ref{fig:rabi_map}. Even the off-resonant oscillations still exhibit a steeper angular characteristic; however, the behaviour diminishes at large detunings. This is consistent with theory, as for a far-off resonant excitation the atoms mostly remain in the $\ket{d}$ state. Overall, the measurement allows us to characterize the system in order to find an optimal working point.

\subsection{Derivation of the two-particle interaction model}\label{sec:two_exc}
 
The energy of two resonantly interacting Rydberg excitations is equal to the energy of a dipole in the electric field of another dipole at the distance $R$:
\begin{equation}
\hat{V}_{\text{dip}}^{\mu  \nu }=\frac{1}{4 \pi  \epsilon _0}\frac{1}{\mathbf{R}^3}\left(\hat{d}_{\mu }\cdot \hat{d}_{\nu }-\frac{3 \left(\hat{d}_{\mu }\cdot \mathbf{R}\right) \left(\hat{d}_{\nu }\cdot \mathbf{R}\right)}{\mathbf{R}^2}\right) 
\end{equation}
For two dipoles coupled with a $\sigma_+$ transition, the energy can be represented in spherical coordinates as:
\begin{equation}
   V_{\text{dip}}^{\mu  \nu }(R,\vartheta) = \frac{C_3}{ R^3}\frac{3 \cos (2 \vartheta )+1}{4}
   \label{eq:pot-met}
\end{equation}
We can consider a pair of Rydberg excitations, one in the mode $d$ and position $\xi$ and the second in the mode $p$ and position $\mu$. 
% We assume that the state of the single Rydberg excitation $\psi$ is symmetric, and thus the two-Rydberg state $\Psi$ is also symmetric and invariant to swap in the position of the excitations.
We assumed that the initial two-Rydberg state $\Psi$ is the product of a single Rydberg excitation $\psi(\mathbf{x}) = \sqrt{p(\mathbf{x})}$, each with a probability density dictated by the normalized atomic distribution $p(\mathbf{x})$ (i.e. the number density of atoms would be equal $Np(\mathbf{x})$). 
Crucially, the two-Rydberg state $\Psi$ is symmetric with respect to swap in the position of the excitations.
Thus, the two-Rydberg-spin-wave wavefunction is:
\begin{equation}
    \Psi\left(\begin{array}{c|c}
        \xi&\multirow{2}{*}{$0$}\\
        \mu&
    \end{array}
    \right) = \psi (\xi) \psi (\mu)
\end{equation}
    And the evolution of the state according to the Schrodinger equation is 
\begin{equation}
    \iu \hbar \dot{\Psi}\left(\begin{array}{c|c}
        \xi&\multirow{2}{*}{$t$}\\
        \mu&
    \end{array}
    \right) = V(\xi-\mu) \Psi\left(\begin{array}{c|c}
        \mu&\multirow{2}{*}{$t$}\\
        \xi&
    \end{array}
    \right) \overset{\text{sym.}}{=} V(\xi-\mu) \Psi\left(\begin{array}{c|c}
        \xi&\multirow{2}{*}{$t$}\\
        \mu&
    \end{array}
    \right)
\end{equation}
Finally, the evolution of the two-Rydberg-spin-wave wavefunction can be found simply as a phase component:
\begin{equation}
    \Psi\left(\begin{array}{c|c}
        \xi&\multirow{2}{*}{$t$}\\
        \mu&
    \end{array}
    \right) = \exp\left(-\frac{\iu t}{\hbar} V(\xi-\mu)\right) \Psi\left(\begin{array}{c|c}
        \xi&\multirow{2}{*}{$0$}\\
        \mu&
    \end{array}
    \right)
    \label{eq:bigpsievol}
\end{equation}
The state of the two-Rydberg excitation is:
\begin{equation}
    \ket{\Psi} = \ket{1,1} = \frac{1}{N}\sum_{i,j} \Psi\left(\begin{array}{c|c}
        \xi_i&\multirow{2}{*}{$0$}\\
        \mu_j&
    \end{array}\right) \frac{\hat{\sigma}^{i}_{dg} \hat{\sigma}^{j}_{pg} \ket{g_1\ldots g_N}}{\sqrt{p(\xi_i) p(\mu_j)}}  
\end{equation}
To characterise the strength of the interactions and the evolution of the state, we calculate the intensity of the electric field of the read-out photons $|\mathcal{E}_d|^2 \propto \langle \hat{d}^\dagger \hat{d} \rangle$ by reading the Rydberg excitations in the mode $d$ from memory with the light-atoms interface. The excitations can be read-out from the memory as photons effectively, only if they are in phase along the entire ensemble. Here we calculate the expected number of excitations in the flat spin-wave mode $d$ for the pair collective state:   

\begin{multline}
    \langle \hat{d}^\dagger \hat{d} \rangle = \frac{1}{N}\sum_{i,j}\bra{\Psi} \hat{\sigma}^{i}_{dg} \hat{\sigma}^{j}_{gd} \ket{\Psi} = \\
    \frac{1}{N^3}\sum_{k,l,m} \Psi^*\left(\begin{array}{c|c}
        \xi_k&\multirow{2}{*}{$t$}\\
        \mu_m&
    \end{array}\right)
    \Psi\left(\begin{array}{c|c}
        \xi_l&\multirow{2}{*}{$t$}\\
        \mu_m&
    \end{array}\right)\frac{1}{\sqrt{p(\xi_k) p(\xi_l)}p(\mu_m)}
\end{multline}
Into this equation, we plug in the result for the evolution (\ref{eq:bigpsievol}). Furthermore, to evaluate the expression, we change summation into integration following a rule $\sum_k\rightarrow\int\mathrm{d}\mathbf{x}Np(\mathbf{x})$, in which the atomic density is an integral measure. We further rewrite the integral, finally obtaining:
\begin{equation}
    \braket{\hat{d}^\dagger \hat{d}} = \int p(\mathbf{x}) \Diff3\mathbf{x} \left|\int p(\mathbf{y}) \exp\left(-\frac{\iu t}{\hbar} V(\mathbf{x}-\mathbf{y})\right) \Diff3\mathbf{y}\right|^2
\end{equation}
Next, we change the variables and assume that $|\mathbf{x}-\mathbf{y}| \ll R$, where $R$ is the diameter of the ensemble, while performing the internal integration. This means that we assume that the density of atoms does not change significantly in the potential range. With this, we get
\begin{multline}
    \braket{\hat{d}^\dagger \hat{d}} = \int p(\mathbf{y}) \Diff3{\mathbf{y}} \left| \int p(\mathbf{x} + \mathbf{y}) \exp \left(-\frac{\iu t}{\hbar} V(\mathbf{x}) \right) \Diff3{\mathbf{x}} \right|^2 \approx \\
    \int p(\mathbf{y}) \Diff3{\mathbf{y}} \biggl| 1 -  p(\mathbf{y})\underbrace{\int \left[1 - \exp \left(-\frac{\iu t}{\hbar} V(\mathbf{x}) \right) \right] \Diff3{\mathbf{x}}}_{A(t)} \biggr|^2 
\end{multline}
The internal integral $A(t)$ can now be performed given the specific potential shape (Eq. \ref{eq:pot-met}), with integration performed in spherical coordinates. Remarkably, we found that the real part of the integral is linear in time, i.e. $\mathfrak{Re}\left[A(t)\right] = Qt$. When expanded for small values of $Qt$, this can be transformed to the simple relation:
\begin{equation}
    \braket{\hat{d}^\dagger \hat{d}} \approx
    \int p(\mathbf{y}) \Diff3{\mathbf{y}} \left[1 - 2 p(\mathbf{y}) Q t\right]
\end{equation}
By integrating the obtained formula with respect to $\mathbf{y}$, where $\int  p^2(\mathbf{y}) \Diff3{\mathbf{y}} =1/\mathcal{V}$, where $\mathcal{V}$ can be interpreted as the ensemble volume, we get:
\begin{equation}
   \braket{\hat{d}^\dagger \hat{d}} \approx 1 - \frac{2 Q}{\mathcal{V}} t \; \text{and} \; Q = \frac{4 \pi^2}{9 \sqrt{3}}  \frac{C_3}{\hbar}
\end{equation}
The factor $Qt$ can be interpreted as volume excluded from the ensemble due to the interactions causing dephasing of spin waves. For only a single atom, it is indeed small compared with the entire volume $\mathcal{V} = \SI{80}{\micro \meter} \times \SI{80}{\micro \meter} \times \SI{4000}{\micro \meter} $, which justifies the applied approximations. The $Q$ itself is the rate at which this volume expands, and interestingly has a dimension of a volumetric flow. Also, we do not consider the evolution for the states $\ket{2,0}$ and $\ket{0,2}$ as there are no dipolar interactions.

\paragraph{Connection between two-particle and toy models}
% \subsection{[krausy 11 połączenie z demko w reżimie $\gamma$-> inf]}
We can now consider the limit of a much stronger interaction (i.e. $\gamma \tau \xrightarrow{} \infty$). In this regime, the mean number of detected photons for the state $\ket{1,1}$ in the model of the interacting pair of Rydberg excitations is $\braket{\hat{d}^\dagger \hat{d}} \approx 0$, as the excluded volume covers the entire cloud. This way, we can describe the evolution with Kraus operators:   
\begin{equation}
\begin{aligned}  
\hat{K}_0 &= \ketbra{0,0}{1,1} \\  
\hat{K}_1 &= \identity - \ketbra{1,1}{1,1} 
\end{aligned}
\end{equation}
Which corresponds to $\braket{\hat{d}^\dagger \hat{d}} = 0$ for the state $\ket{1,1}$ and no changes for other states. This reproduces the Kraus operators proposed in the toy model.

\subsection{Generalization to multi-particle interaction model} \label{sec:phenomenological}
\paragraph{Interaction-induced decay}
On the other hand, in the regime of weak interactions, we can extrapolate the result of two interacting excitations to the regime of many ($n_p$) control atoms, each generating such an exclusion volume.  Due to the small number of initial excitations compared to the number of atoms, we can assume that there are only interactions between pairs of Rydberg excitations, and we can neglect many-body interactions. It is safe to assume that an interaction with multiple control atoms cannot restore the phase-matched readout. We can thus assume that each control atom acts independently. For large $n_p$ and small excluded volume of any one atom, we have:
\begin{equation}
      \braket{\hat{d}^\dagger \hat{d}} \approx \left(1 - \frac{2 Q }{\mathcal{V}} t\right)^{n_p} \approx \exp\left({-n_p\frac{2Q}{\mathcal{V}} t}\right)
\end{equation}
This, of course, takes into account that exclusion volumes can overlap, and for large time $t$, even the entire ensemble may be dephased.
For the states chosen in the experiment we have $\frac{C_3}{2\pi\hbar} = \SI{3.709}{\GHz \: \um^3}$ \cite{SIBALIC2017319}. As indicated in the main text, we define a per-atom decoherence rate (Eq. \ref{eq:gammath}), which agrees remarkably well with an experimental estimate. Thus, the observation of this decay rate as well as the profile of the decay being exponential (Fig. \ref{fig:exp_data_all}(b)) strongly suggests an excellent viability of the proposed model.

\paragraph{Excitation number distribution}
To fully understand the experimental situation, we need a self-consistent treatment of the statistical properties of the system. We achieve this by introducing a multi-particle model, where each mode decays proportionally to the occupation of the other mode, which results in the behaviour given by Eqs. (\ref{eq:operator_evolution}).
To describe the evolution of the state under the influence of this operation, we can define the asymmetric Kraus operator for the single mode, similarly to the beam splitter \cite{Leviant2022quantumcapacity} with losses regulated by the opposite mode:
\begin{equation}
    \hat{K}_l =  \hat{d}^l \sqrt{\frac{(e^{\gamma \tau \hat{p}^\dagger \hat{p}}-1)^l}{l!}}e^{-\gamma \tau \hat{p}^\dagger \hat{p} \hat{d}^\dagger \hat{d}/2}
\end{equation}
To calculate the FI, we will need the probability distribution of the number of detected photons. We can find it by considering the evolution of the initial bi-coherent state $\ket{\psi} = \ket{\delta,\beta}$, which is a tensor product of two coherent states with amplitudes $\delta$ and $\beta$ in modes $d$ and $p$, respectively. This way, we obtain the probability distribution of the number of detected photons in the mode $d$:
\begin{multline}
    P(n_d) = \sum_{l,n_p} |\bra{n_d,n_p} \hat{K}_l \ket{\psi}|^2 = \\
    = \sum_{n_p} \frac{|\beta|^{2 n_p} \left(|\delta|^2 e^{- \gamma \tau  n_p}\right)^{n_d} e^{-|\delta|^2 e^{- \gamma \tau  n_p}-|\beta|^2}}{n_p! n_d!}
\end{multline}

From the photon number distribution, we can calculate the mean number of detected photons in the initial Rydberg state:
\begin{equation}
    \langle n_d' \rangle=\bra{\delta,\beta}\hat{d}'^\dagger\hat{d}'\ket{\delta,\beta} = |\delta|^2\exp{\left[-|\beta|^2(1-e^{-\gamma \tau})\right]} 
\end{equation}
Note that the value of the mean number of photons is almost identical to the one obtained in the two-interacting-excitation model. The difference in the multi-particle model comes from considering coherent states. The probability distribution resembles a Poissonian distribution affected by random losses, the strength of which is also Poisson-distributed. 

In the experiment, we rotate the state with the MW field, which transforms the coherent states amplitudes as functions of rotation angle $\theta$ and incident mean number of photons $\Bar{n}$ ($|\delta|^2 \rightarrow \Bar{n}\cos^2(\theta/2), |\beta|^2 \rightarrow \Bar{n}\sin^2(\theta/2)$). Now with the unravelled $\theta$ dependence, the probability distribution $P_\theta (n_d)$ can be used to calculate the classical FI for a detected photon as in Eq.~\eqref{eq:fisher_information}, and we can utilise the Cramér-Rao bound to get the ultimate precision of the estimation.

\begin{figure}[t]
\centering
\includegraphics[width = 1\columnwidth]{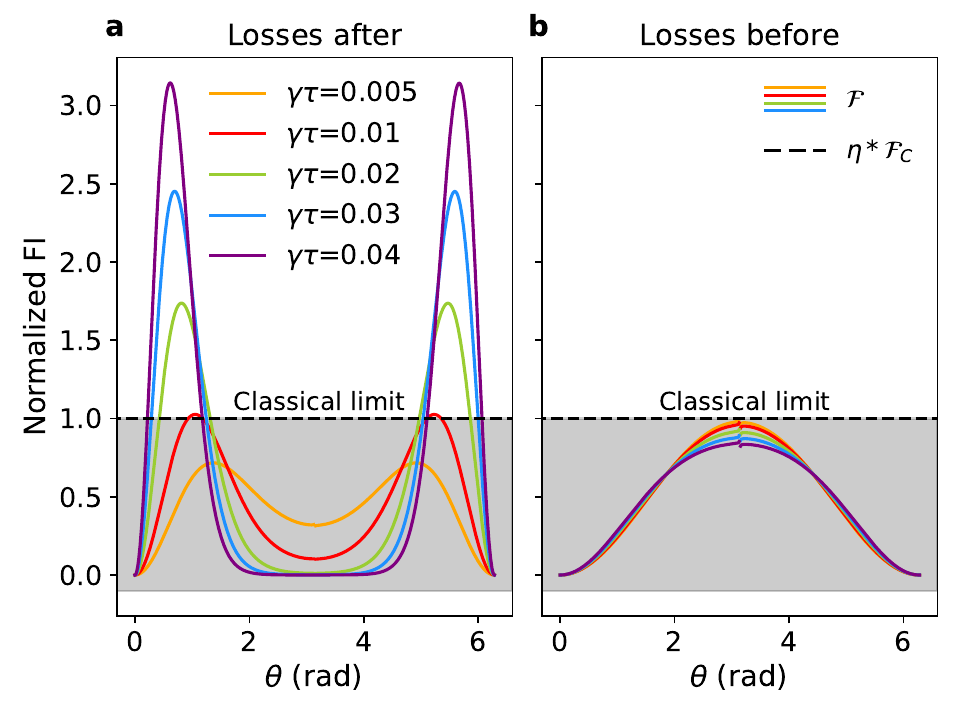}
\caption{\label{fig:fisher_num}(Extended Data Figure) Plot of Fisher information for different values of the parameter $\gamma$ for efficiency $\eta = 0.02$ after the interaction\subfig{a} and before the interactions \subfig{b}. Fisher information at the beginning of the experiment without loss is $\mathcal{F}_C = 1$. In the case where losses are introduced after the interactions, there is an improvement in the Fisher information depending on the value of the decay rate $\gamma$. The Fisher information on both plots is divided by $\eta$.}
\end{figure}

The probability distribution $P_\theta(n_d)$ allows us to conduct the numerical simulations of the precision of the estimation for the different values of the parameter $\gamma$. To showcase the robustness of this framework, we give a simple example. In Fig.~\ref{fig:fisher_num}\subfig{a}, the losses are inflicted after the Rydberg interactions, showing the metrological advantage over the classical value of the FI with the losses, just as in the main text of the article.  In Fig.~\ref{fig:fisher_num}\subfig{b}, the losses are applied before the interactions, displaying the lack of advantage over the procedure without the interactions. 
The classical limit for the estimation precision is the initial FI of the coherent state multiplied by efficiency $\eta$. 

As the described operations of regulated loss acts in the same way for either of the modes, to fully describe the evolution of the state in both modes, we can symmetrise the Kraus operator. The operation $\Lambda$, in this case, could also be represented as a set $\cup_{k,m}\{\hat{K}_{k,m}\}$ of Kraus operators:
\begin{equation}\label{eq:kraus_lambda}
    \hat{K}_{k,m} = \hat{p}^k \hat{d}^ m \sqrt{\frac{(e^{\gamma\tau \hat{d}^\dagger \hat{d}}-1)^k(e^{\gamma\tau\hat{p}^\dagger \hat{p}}-1)^m}{k!m!}} e^{-\gamma\tau\hat{d}^\dagger \hat{d} \hat{p}^\dagger \hat{p} } 
\end{equation}
This set of operators allows us to describe the evolution of the states in both modes.

\subsection{Connection between toy and multi-particle models}\label{sec:conn}
We can now revisit the limit of stronger interactions. 
Assume that we are working in a Hilbert space limited to at most two bosonic excitations in two modes (the same as in the toy model). We can represent the Kraus operators given in Eq.~\eqref{eq:kraus_lambda} in the Fock basis of the Hilbert space. In the limit $\gamma \tau \xrightarrow{} \infty$ the only non-zero operators are:
\begin{subequations}
\begin{gather}
\hat{K}_{1,1} = \ketbra{0,0}{1,1}\\
\begin{split}
\hat{K}_{0,0} &= \ketbra{2,0}{2,0} + \ketbra{1,0}{1,0} + \ketbra{0,2}{0,2}\\
        &+ \ketbra{0,1}{0,1} + \ketbra{0,0}{0,0}\\
        &= \identity - \ketbra{1,1}{1,1}
\end{split}
\end{gather}
\end{subequations}
We recover the known operators from the toy model, showing the backwards compatibility. This means that for at most two excitations, the interaction is optimal in the limit of large $\gamma \tau$.

\subsection{Derivation of the FI for the toy model}
Here we consider the two-particle toy model introduced in the main text,
where the parameter $\theta$ is encoded in the state $\ket{\psi_\theta}$ given in \eqref{rotated20state}.

In the absence of losses, the ideal  excitations measurement would give us one of three outcomes $(2,0)$, $(1,1)$, $(0,2)$ with probabilities corresponding to the squared state amplitudes in Eq.~\eqref{rotated20state}.
In the presence of losses, we need to include three other possible measurement outcomes $(1,0)$, $(0,1)$,
$(0,0)$ representing one surviving excitation in one of the two modes, or no observed excitation at all.

In order to model this situation mathematically, let us introduce six generalised measurement operators $M_{(i,j)}$ acting on $\mathcal{H}=\textrm{span}\left(\ket{2,0}, \ket{1,1}, \ket{0,2}\right)$: 
\begin{equation}
\label{eq:povmslosses}
\begin{aligned}
    M_{(2,0)} &= \eta^2 \proj{2,0}, \ M_{(0,2)} = \eta^2 \proj{0,2}, \\
    M_{(1,1)} &= \eta^2 \proj{1,1}, \ M_{(0,0)} = (1-\eta)^2 \openone, \\ M_{(1,0)}&=2\eta(1-\eta) \proj{2,0} + \eta(1-\eta)\proj{1,1}, \\
     M_{(0,1)}&=2\eta(1-\eta) \proj{0,2} + \eta(1-\eta)\proj{1,1},
\end{aligned}
\end{equation}
and the corresponding probabilities of observing a given event $(i,j)$ should be computed via:
\begin{equation}
    p_{(i,j)} = \bra{\psi_\theta} M_{(i,j)} \ket{\psi_\theta}.
\end{equation}
Using the classical FI formula, we can compute the corresponding FI in a straightforward way, arriving at:
\begin{equation}
\label{eq:filossnoec}
\mathcal{F} = \sum_{(i,j)} \frac{1}{p_{(i,j)}} \left(\frac{\textrm{d} p_{(i,j)}}{\textrm{d} \theta}\right)^2 = 2 \eta,
\end{equation}
which indicates the typical FI reduction due to loss  for independent probes.

Let us now consider the error-prevention protocol 
described by Kraus operators $\hat{K}_0 = 
\ketbra{0,0}{1,1}$ and $\hat{K}_1 = \openone - \ketbra{1,1}{1,1}$ that effectively transfer the $\ket{1,1}$ 
state to $\ket{0,0}$. 
This operation effectively prepares a mixed state of the form:
\begin{equation}
    \tilde{\rho}_\theta = \proj{\tilde{\psi}_\theta} + \frac{1}{2}\sin^2\theta \,\proj{0,0},
\end{equation}
where unnormalized state $\ket{\tilde{\psi}_\theta}$ reads
\begin{equation}
\ket{\tilde{\psi}_\theta} = \cos^2\left(\frac{\theta}{2}\right) \ket{2,0} - \sin^2\left(\frac{\theta}{2}\right) \ket{0,2}
\end{equation}
and the last term represents the part that was transferred from the $\ket{1,1}$ component. 

We may now compute the detection probabilities for the lossy measurement using
\begin{equation}
\tilde{p}_{(i,j)} =  \textrm{Tr}\left(\tilde{\rho}_\theta  \tilde{M}_{(i,j)} \right),
\end{equation}
where $\tilde{M}_{(i,j}$ are identical to the measurement operators given in \eqref{eq:povmslosses} treated as operators acting on the space $\tilde{\mathcal{H}} = \textrm{span}(\ket{2,0}, \ket{0,2}, \ket{0,0})$ (we simply ignore the $\ket{1,1}$ projectors)
with the only modification for 
\begin{equation}
\tilde{M}_{(0,0)} = (1-\eta)^2\left(\proj{2,0} + \proj{0,2}\right) + \proj{0,0},
\end{equation}
which now also includes the contribution from the $\ket{0,0}$ state.

Applying the general FI formula \eqref{eq:filossnoec} to 
$\tilde{p}_{(i,j)}$ we get:
\begin{equation}
\tilde{\mathcal{F}} = \frac{2 \sin^2 \theta \left[1 +(1-\eta)^2\right]}{\eta \sin^2 \theta + (1-\eta)^2/(1-\eta/2)}  \overset{\theta=\pi/2}{=} 2\eta(2-\eta),
\end{equation}
which achieves the maximal value for $\theta=\pi/2 + k\pi$,
as depicted in Fig.~\ref{fig:sim}\subfig{d}.

\subsection{Optimality of the pre-measurement processing protocol}
One may wonder if the applied protocol is optimal.
To answer this question, we considered  a general quantum operation $\Lambda$  on the sensing system (general completely positive map \cite{Nielsen2000}), that acts after the parameter encoding stage $U(\theta)$ and before the state experience detection losses $\Xi(\eta)$. Let us define the \emph{ultimate} QFI as:  
\begin{equation}
 \mathcal{F}_Q^\diamond = \max_{\Lambda} \mathcal{F}_Q\left[\Xi(\eta) \circ \Lambda \circ U(\theta) \ket{\psi_0} \right],
\end{equation}
which represents the maximal QFI that we can achieve if we optimize the scheme over a general error-correcting operation $\Lambda$ on the system---note that formally here, apart from including the detection losses, we do not assume any particular measurement being performed in the end, which makes the approach slightly more general than the one adopted in \cite{Len2022-vt, PRXQuantum.4.040305}.  

The above optimisation problem can be solved via an iterative optimisation procedure described in \cite{Kurdzialek2025} and implemented in the QMetro++ package \cite{pdulian2025pdulian} that allows optimising the QFI of the output state, over quantum channels representing a general estimation strategy. 
In the present context, we have applied the tools to optimise the pre-processing operation $\Lambda$, with input probe state fixed and no additional ancillary systems allowed. Thanks to the relatively small dimension of the problem considered (states with at most two total excitations in both modes form a 6-dimensional Hilbert space), the optimisation is very efficient. Remarkably, the performance of the numerically found optimal solution coincides perfectly with the simple error-prevention scheme described above. We deposited the optimisation script along with the data used for plots at \cite{MJMXI8_2025}.

To strengthen the above 'numerical proof', we may also use analytical upper bounds from \cite{PRXQuantum.4.040305}. 
More specifically, we may apply results of Sec.~IVD from \cite{PRXQuantum.4.040305} that deal with a pure state case and commuting measurement operators, which is exactly the situation we are considering.  There, it has been shown that whatever the error-prevention operation $\Lambda$ is applied, one may not obtain a larger FI than 
\begin{equation}
\tilde{\mathcal{F}} \leq \gamma \mathcal{F}_Q = 2 \gamma,
\end{equation}
where $\gamma$ is a coefficient that may be computed 
using equation (64-67) from \cite{PRXQuantum.4.040305}. Since the bound in \cite{PRXQuantum.4.040305} formally requires the measurement operators to be strictly positive, one needs to introduce some regularization of the operators by an infinitesimal $\epsilon$ parameter guaranteeing positivity, for which in the  end will take the limit $\epsilon \rightarrow 0$. 
After some lengthy but straightforward calculations, this procedure leads us to the desired solution:
\begin{equation}
\gamma = \eta(2 -\eta),
\end{equation}
proving the optimality of the protocol.

\end{document}